\documentclass{PoS}
\usepackage{latexsym}
\usepackage{amsmath}

\title{Strong to weak coupling transitions of SU(N) gauge theories in 2+1 dimensions}

\ShortTitle{Strong to weak coupling transitions of SU(N) gauge theories in 2+1 dimensions}

\author{\speaker{Francis Bursa} and Michael Teper\\

        Rudolf Peierls Centre for Theoretical Physics, University of Oxford,\\
	1 Keble Road, Oxford OX1 3NP, U.K.\\

        E-mail: \email{f.bursa1@physics.ox.ac.uk}}

\abstract{We find a strong-to-weak coupling cross-over in $D=2+1$ $SU(N)$ 
lattice gauge theories
that appears to become a third-order phase transition at $N=\infty$,
in a similar way to the Gross-Witten 
transition in the $D=1+1$ $SU(N\to\infty)$ lattice gauge theory. 
There is, in addition, a peak in the specific heat at approximately the same coupling
that increases with $N$, which is connected to $Z_N$ monopoles 
(instantons), reminiscent of the first order bulk transition 
that occurs in $D=3+1$ for $N\geq 5$.
Our calculations are not precise enough to determine whether this peak
is due to a second-order phase transition at $N=\infty$ or to
a third-order phase transition with different critical behaviour
to that of the Gross-Witten transition. We investigate whether the trace of the Wilson loop has a 
non-analyticity in the coupling at some critical area, but find 
no evidence for this. 
However we do find that, just as one can prove occurs in $D=1+1$, 
the eigenvalue density of a Wilson loop forms a gap at $N=\infty$ 
at a critical value of its trace. We show that
this gap formation is in fact a corollary of a remarkable
similarity between the eigenvalue spectra of Wilson loops
in $D=1+1$ and $D=2+1$ (and indeed $D=3+1$): for the same value
of the trace, the eigenvalue spectra are nearly identical.
This holds for finite as well as infinite $N$; irrespective
of the Wilson loop size in lattice units;
and for Polyakov as well as Wilson loops.}

\FullConference{XXIVth International Symposium on Lattice Field Theory\\

                July 23-28, 2006\\

                Tucson, Arizona, USA}

\begin{document}

\section{Introduction}
A phase transition requires an infinite number of degrees of freedom.
One can have phase transitions
on finite volumes at $N=\infty$ because in
this case we have an infinite number of degrees of freedom at each 
point in space. The classic example in the context of gauge field
theories is the Gross-Witten
transition 
\cite{GW}
that occurs in the $D=1+1$ $SU(\infty)$ lattice gauge theory.

In $D=3+1$ $SU(N)$ gauge theories numerical studies reveal the
existence for $N\geq 5$ of a first order `bulk' transition separating 
the weak and strong coupling regions 
\cite{OxG01,OxT05}.

These are both 
in some sense strong to weak coupling
transitions which has led to the conjecture
\cite{NNect04,NNlat05}
that Wilson loops in general will show such $N=\infty$ transitions
when the physical size of the
loop passes some critical value. 
Such a transition in $D=3+1$ could
have interesting implications for dual string approaches to
large-$N$ gauge theories and provide a natural 
explanation for the observed rapid onset of
non-perturbative physics in the strong interactions
\cite{GW,MTect04,NNlat05}.


\section{Background}
\subsection{The `Gross-Witten' transition}

The $D=1+1$ $SU(N)$ lattice
gauge theory can be explicitly solved~\cite{GW}. One finds a 
cross-over between weak and strong coupling that sharpens
with increasing $N$ into a third order phase transition at
$N=\infty$. In terms of the plaquette $u_p$ 
this shows up in a change of functional behaviour
\begin{equation}
\langle u_p\rangle \stackrel{N\to\infty}{=} \begin{cases}
\frac{1}{\lambda}  & \lambda\geq 2, \\
1 - \frac{\lambda}{4}  &  \lambda\leq 2.
\end{cases}
\label{eqn_upGW}
\end{equation}
More detailed information about the behaviour of plaquettes and 
Wilson loops can be gained by considering their eigenvalues, which are
are just phases, $\lambda = \exp\{i\alpha\}$.
As $\beta\to 0$ the eigenvalue distribution $\rho (\alpha )$ of 
a Wilson loop becomes uniform while as  $\beta\to\infty$ it 
becomes increasingly peaked around $\alpha=0$.
At the Gross--Witten transition a gap opens in the density
of plaquette eigenvalues: in the strongly--coupled phase
the eigenvalue density is non--zero for all angles,
but in the weakly-coupled phase 
it is only non--zero in the range 
$-\alpha_c \leq \alpha \leq \alpha_c$, where $\alpha_c < \pi$
\cite{GW}. 

In $D=3+1$ it is known that for $N\geq 5$
\cite{OxG01,OxT05}
there is a strong first order transition as $\beta$ is varied 
from strong to weak coupling. Calculations in progress
\cite{BurTepVai}
suggest that the plaquette eigenvalue distribution shows a gap formation at $N=\infty$ that is similar to the
$D=1+1$ transition.

In $D=2+1$ there has been, as far as we are aware, no systematic
search for a Gross-Witten or `bulk' transition, and this is one
of the gaps that the present work intends to fill.

\subsection{Wilson loop transitions}
The Gross-Witten transition involves the smallest possible
Wilson loop, the plaquette. On the weak coupling side the plaquette
can be calculated in perturbation
theory; but this breaks down abruptly at the Gross-Witten transition~\cite{GW}.
The coupling is the bare coupling and hence a coupling on the
length scale of the plaquette. One might interpret this
as saying that there is a 
critical length scale at which perturbation theory in the running coupling
will suddenly break down. 

One might imagine that this generalises to other Wilson loops: i.e.
when we scale up a Wilson loop, at some critical size, in 
`physical units', there is a non-analyticity. In fact precisely
such a scenario has been conjectured for $SU(N\to\infty)$ gauge
theories in $D=3+1$
\cite{NNect04,NNlat05}.

Such a  non-analyticity does in fact occur for the $SU(N\to\infty)$
continuum theory in $D=1+1$
\cite{DurOle,BasGriVian}. 
The transition occurs at a fixed physical area 
$A_{crit}=\frac{8}{g^2N}$.
Very much larger Wilson loops have a flat 
eigenvalue spectrum $\rho(\alpha)$ which becomes peaked  
as $A \to A^{+}_{crit}$. As   $A$  decreases through $A_{crit}$
a gap appears in the spectrum near the extreme phases 
$\alpha = \pm \pi$. However, unlike the Gross-Witten transition
this is not a phase transition: the partition function
is analytic. Thus it is unclear what if any
is the physical significance of this non-analyticity.

In this talk we investigate whether such a non-analyticity 
develops in  $D=2+1$.

\section{Results}
\subsection{Preliminaries}

At a phase transition appropriate derivatives of $\frac{1}{V}\log Z$,
where $V$ is the volume and  $Z$ is the partition function, will 
diverge or be discontinuous as $V\to\infty$. The lowest order of such 
a singular derivative determines the order of the phase transition.

With the standard plaquette action, a conventional first order 
transition has a discontinuity at $V = \infty$ in the average 
plaquette.

A conventional second order transition has a continuous first 
derivative of $Z$  but a diverging second derivative and a specific heat
$C \to \infty$ as $V\to\infty$. Defining $\overline{u_p}$ to be the average  
value of $u_p$ over the space-time volume for a single lattice
field, $C$ can be written as
\begin{equation}
C = 
N_p(\langle \overline{u_p}^2\rangle-\langle \overline{u_p} \rangle^2).
\label{eqn_C}
\end{equation}

A conventional third order transition has continuous first and second 
order derivatives but a singular third-order derivative,
$C^{\prime}\equiv N^{-1}_p \partial^3 \log Z/\partial\beta^3$,
at $V =\infty$. This may be written as
\begin{equation}
C^\prime =
N_p^2 (\langle\overline{u_p}^3\rangle
-3\langle\overline{u_p}\rangle\langle\overline{u_p}^2\rangle
+2\langle\overline{u_p}\rangle^3).
\label{eqn_Cprime}
\end{equation}

Since, in general, fluctuations 
in the pure gauge theory decrease by powers of $N$
we define the rescaled quantities
$C_2=N^2\times C$ and $C_3=N^4 \times C^\prime$
which one expects generically to have finite non-zero limits
when $N\to\infty$. If we find a crossover in $C_2$ or $C_3$ which 
does not sharpen with increasing volume at fixed $N$, but 
rather becomes a divergence or a discontinuity 
only in the large--$N$ limit, then this will indicate a 
second-- or third--order $N=\infty$ phase transition 
respectively.

Since large-$N$ phase transitions can arise from completely local fluctuations
we also consider local versions 
of $C_2$ and $C_3$ where we replace $\overline{u_p}$
by $u_p$, which we call $P_2$ and $P_3$ respectively.

To search for non-analyticities in Wilson loops we search for non-analyticities in  $\langle u_w \rangle$
and its derivatives. We 
also calculate `local' versions of the latter,
just as we do for $\langle u_p \rangle$, and various
moments of the Wilson loops. Finally, we also
calculate and analyse their eigenvalue spectra.

\subsection{Bulk transition}
In 3+1 dimensions the bulk transition is easily visible as a large 
discontinuity in the action for $N\geq 5$ and as a (finite) peak in the specific heat for
$N\leq 4$. We have searched 
for an analogous jump or rapid crossover in 2+1 dimensional
$SU(6)$, $SU(12)$, $SU(24)$ and $SU(48)$ gauge theories. What we see
is that the action appears to be approaching 
a smooth crossover in the large--$N$ limit.

Our results for the specific heat for $SU(6)$ and $SU(12)$ show a clear peak around $\gamma\simeq 0.42$ which appears to grow stronger with increasing $N$.
We have calculated $P_2$, the `local' version of $C_2$, for $SU(6)$, $SU(12)$, $SU(24)$ and $SU(48)$. We see no significant evidence for a peak, which indicates that if there is a second order transition 
at $N=\infty$ it will primarily
involve correlations between different plaquettes rather than
arising from the fluctuations of individual plaquettes.
However, what we do see in $P_2$ 
is definite evidence for a cusp developing at $\gamma \simeq 0.43$.

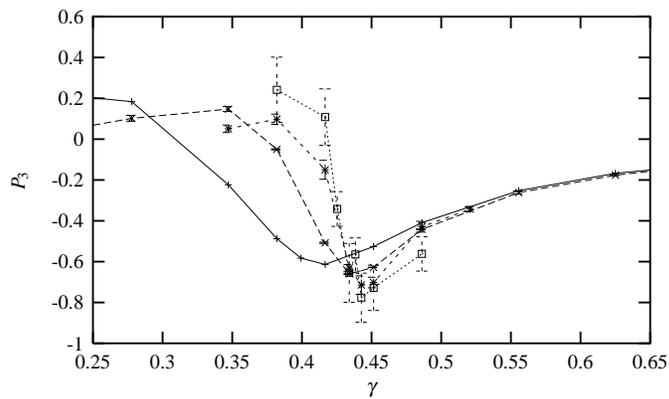
\begin	{figure}[p]
\begin	{center}
\leavevmode
\scalebox{0.7}{
\begingroup%
  \makeatletter%
  \newcommand{\GNUPLOTspecial}{%
    \@sanitize\catcode`\%=14\relax\special}%
  \setlength{\unitlength}{0.1bp}%
{\GNUPLOTspecial{!
/gnudict 256 dict def
gnudict begin
/Color false def
/Solid false def
/gnulinewidth 5.000 def
/userlinewidth gnulinewidth def
/vshift -33 def
/dl {10 mul} def
/hpt_ 31.5 def
/vpt_ 31.5 def
/hpt hpt_ def
/vpt vpt_ def
/M {moveto} bind def
/L {lineto} bind def
/R {rmoveto} bind def
/V {rlineto} bind def
/vpt2 vpt 2 mul def
/hpt2 hpt 2 mul def
/Lshow { currentpoint stroke M
  0 vshift R show } def
/Rshow { currentpoint stroke M
  dup stringwidth pop neg vshift R show } def
/Cshow { currentpoint stroke M
  dup stringwidth pop -2 div vshift R show } def
/UP { dup vpt_ mul /vpt exch def hpt_ mul /hpt exch def
  /hpt2 hpt 2 mul def /vpt2 vpt 2 mul def } def
/DL { Color {setrgbcolor Solid {pop []} if 0 setdash }
 {pop pop pop Solid {pop []} if 0 setdash} ifelse } def
/BL { stroke userlinewidth 2 mul setlinewidth } def
/AL { stroke userlinewidth 2 div setlinewidth } def
/UL { dup gnulinewidth mul /userlinewidth exch def
      10 mul /udl exch def } def
/PL { stroke userlinewidth setlinewidth } def
/LTb { BL [] 0 0 0 DL } def
/LTa { AL [1 udl mul 2 udl mul] 0 setdash 0 0 0 setrgbcolor } def
/LT0 { PL [] 1 0 0 DL } def
/LT1 { PL [4 dl 2 dl] 0 1 0 DL } def
/LT2 { PL [2 dl 3 dl] 0 0 1 DL } def
/LT3 { PL [1 dl 1.5 dl] 1 0 1 DL } def
/LT4 { PL [5 dl 2 dl 1 dl 2 dl] 0 1 1 DL } def
/LT5 { PL [4 dl 3 dl 1 dl 3 dl] 1 1 0 DL } def
/LT6 { PL [2 dl 2 dl 2 dl 4 dl] 0 0 0 DL } def
/LT7 { PL [2 dl 2 dl 2 dl 2 dl 2 dl 4 dl] 1 0.3 0 DL } def
/LT8 { PL [2 dl 2 dl 2 dl 2 dl 2 dl 2 dl 2 dl 4 dl] 0.5 0.5 0.5 DL } def
/Pnt { stroke [] 0 setdash
   gsave 1 setlinecap M 0 0 V stroke grestore } def
/Dia { stroke [] 0 setdash 2 copy vpt add M
  hpt neg vpt neg V hpt vpt neg V
  hpt vpt V hpt neg vpt V closepath stroke
  Pnt } def
/Pls { stroke [] 0 setdash vpt sub M 0 vpt2 V
  currentpoint stroke M
  hpt neg vpt neg R hpt2 0 V stroke
  } def
/Box { stroke [] 0 setdash 2 copy exch hpt sub exch vpt add M
  0 vpt2 neg V hpt2 0 V 0 vpt2 V
  hpt2 neg 0 V closepath stroke
  Pnt } def
/Crs { stroke [] 0 setdash exch hpt sub exch vpt add M
  hpt2 vpt2 neg V currentpoint stroke M
  hpt2 neg 0 R hpt2 vpt2 V stroke } def
/TriU { stroke [] 0 setdash 2 copy vpt 1.12 mul add M
  hpt neg vpt -1.62 mul V
  hpt 2 mul 0 V
  hpt neg vpt 1.62 mul V closepath stroke
  Pnt  } def
/Star { 2 copy Pls Crs } def
/BoxF { stroke [] 0 setdash exch hpt sub exch vpt add M
  0 vpt2 neg V  hpt2 0 V  0 vpt2 V
  hpt2 neg 0 V  closepath fill } def
/TriUF { stroke [] 0 setdash vpt 1.12 mul add M
  hpt neg vpt -1.62 mul V
  hpt 2 mul 0 V
  hpt neg vpt 1.62 mul V closepath fill } def
/TriD { stroke [] 0 setdash 2 copy vpt 1.12 mul sub M
  hpt neg vpt 1.62 mul V
  hpt 2 mul 0 V
  hpt neg vpt -1.62 mul V closepath stroke
  Pnt  } def
/TriDF { stroke [] 0 setdash vpt 1.12 mul sub M
  hpt neg vpt 1.62 mul V
  hpt 2 mul 0 V
  hpt neg vpt -1.62 mul V closepath fill} def
/DiaF { stroke [] 0 setdash vpt add M
  hpt neg vpt neg V hpt vpt neg V
  hpt vpt V hpt neg vpt V closepath fill } def
/Pent { stroke [] 0 setdash 2 copy gsave
  translate 0 hpt M 4 {72 rotate 0 hpt L} repeat
  closepath stroke grestore Pnt } def
/PentF { stroke [] 0 setdash gsave
  translate 0 hpt M 4 {72 rotate 0 hpt L} repeat
  closepath fill grestore } def
/Circle { stroke [] 0 setdash 2 copy
  hpt 0 360 arc stroke Pnt } def
/CircleF { stroke [] 0 setdash hpt 0 360 arc fill } def
/C0 { BL [] 0 setdash 2 copy moveto vpt 90 450  arc } bind def
/C1 { BL [] 0 setdash 2 copy        moveto
       2 copy  vpt 0 90 arc closepath fill
               vpt 0 360 arc closepath } bind def
/C2 { BL [] 0 setdash 2 copy moveto
       2 copy  vpt 90 180 arc closepath fill
               vpt 0 360 arc closepath } bind def
/C3 { BL [] 0 setdash 2 copy moveto
       2 copy  vpt 0 180 arc closepath fill
               vpt 0 360 arc closepath } bind def
/C4 { BL [] 0 setdash 2 copy moveto
       2 copy  vpt 180 270 arc closepath fill
               vpt 0 360 arc closepath } bind def
/C5 { BL [] 0 setdash 2 copy moveto
       2 copy  vpt 0 90 arc
       2 copy moveto
       2 copy  vpt 180 270 arc closepath fill
               vpt 0 360 arc } bind def
/C6 { BL [] 0 setdash 2 copy moveto
      2 copy  vpt 90 270 arc closepath fill
              vpt 0 360 arc closepath } bind def
/C7 { BL [] 0 setdash 2 copy moveto
      2 copy  vpt 0 270 arc closepath fill
              vpt 0 360 arc closepath } bind def
/C8 { BL [] 0 setdash 2 copy moveto
      2 copy vpt 270 360 arc closepath fill
              vpt 0 360 arc closepath } bind def
/C9 { BL [] 0 setdash 2 copy moveto
      2 copy  vpt 270 450 arc closepath fill
              vpt 0 360 arc closepath } bind def
/C10 { BL [] 0 setdash 2 copy 2 copy moveto vpt 270 360 arc closepath fill
       2 copy moveto
       2 copy vpt 90 180 arc closepath fill
               vpt 0 360 arc closepath } bind def
/C11 { BL [] 0 setdash 2 copy moveto
       2 copy  vpt 0 180 arc closepath fill
       2 copy moveto
       2 copy  vpt 270 360 arc closepath fill
               vpt 0 360 arc closepath } bind def
/C12 { BL [] 0 setdash 2 copy moveto
       2 copy  vpt 180 360 arc closepath fill
               vpt 0 360 arc closepath } bind def
/C13 { BL [] 0 setdash  2 copy moveto
       2 copy  vpt 0 90 arc closepath fill
       2 copy moveto
       2 copy  vpt 180 360 arc closepath fill
               vpt 0 360 arc closepath } bind def
/C14 { BL [] 0 setdash 2 copy moveto
       2 copy  vpt 90 360 arc closepath fill
               vpt 0 360 arc } bind def
/C15 { BL [] 0 setdash 2 copy vpt 0 360 arc closepath fill
               vpt 0 360 arc closepath } bind def
/Rec   { newpath 4 2 roll moveto 1 index 0 rlineto 0 exch rlineto
       neg 0 rlineto closepath } bind def
/Square { dup Rec } bind def
/Bsquare { vpt sub exch vpt sub exch vpt2 Square } bind def
/S0 { BL [] 0 setdash 2 copy moveto 0 vpt rlineto BL Bsquare } bind def
/S1 { BL [] 0 setdash 2 copy vpt Square fill Bsquare } bind def
/S2 { BL [] 0 setdash 2 copy exch vpt sub exch vpt Square fill Bsquare } bind def
/S3 { BL [] 0 setdash 2 copy exch vpt sub exch vpt2 vpt Rec fill Bsquare } bind def
/S4 { BL [] 0 setdash 2 copy exch vpt sub exch vpt sub vpt Square fill Bsquare } bind def
/S5 { BL [] 0 setdash 2 copy 2 copy vpt Square fill
       exch vpt sub exch vpt sub vpt Square fill Bsquare } bind def
/S6 { BL [] 0 setdash 2 copy exch vpt sub exch vpt sub vpt vpt2 Rec fill Bsquare } bind def
/S7 { BL [] 0 setdash 2 copy exch vpt sub exch vpt sub vpt vpt2 Rec fill
       2 copy vpt Square fill
       Bsquare } bind def
/S8 { BL [] 0 setdash 2 copy vpt sub vpt Square fill Bsquare } bind def
/S9 { BL [] 0 setdash 2 copy vpt sub vpt vpt2 Rec fill Bsquare } bind def
/S10 { BL [] 0 setdash 2 copy vpt sub vpt Square fill 2 copy exch vpt sub exch vpt Square fill
       Bsquare } bind def
/S11 { BL [] 0 setdash 2 copy vpt sub vpt Square fill 2 copy exch vpt sub exch vpt2 vpt Rec fill
       Bsquare } bind def
/S12 { BL [] 0 setdash 2 copy exch vpt sub exch vpt sub vpt2 vpt Rec fill Bsquare } bind def
/S13 { BL [] 0 setdash 2 copy exch vpt sub exch vpt sub vpt2 vpt Rec fill
       2 copy vpt Square fill Bsquare } bind def
/S14 { BL [] 0 setdash 2 copy exch vpt sub exch vpt sub vpt2 vpt Rec fill
       2 copy exch vpt sub exch vpt Square fill Bsquare } bind def
/S15 { BL [] 0 setdash 2 copy Bsquare fill Bsquare } bind def
/D0 { gsave translate 45 rotate 0 0 S0 stroke grestore } bind def
/D1 { gsave translate 45 rotate 0 0 S1 stroke grestore } bind def
/D2 { gsave translate 45 rotate 0 0 S2 stroke grestore } bind def
/D3 { gsave translate 45 rotate 0 0 S3 stroke grestore } bind def
/D4 { gsave translate 45 rotate 0 0 S4 stroke grestore } bind def
/D5 { gsave translate 45 rotate 0 0 S5 stroke grestore } bind def
/D6 { gsave translate 45 rotate 0 0 S6 stroke grestore } bind def
/D7 { gsave translate 45 rotate 0 0 S7 stroke grestore } bind def
/D8 { gsave translate 45 rotate 0 0 S8 stroke grestore } bind def
/D9 { gsave translate 45 rotate 0 0 S9 stroke grestore } bind def
/D10 { gsave translate 45 rotate 0 0 S10 stroke grestore } bind def
/D11 { gsave translate 45 rotate 0 0 S11 stroke grestore } bind def
/D12 { gsave translate 45 rotate 0 0 S12 stroke grestore } bind def
/D13 { gsave translate 45 rotate 0 0 S13 stroke grestore } bind def
/D14 { gsave translate 45 rotate 0 0 S14 stroke grestore } bind def
/D15 { gsave translate 45 rotate 0 0 S15 stroke grestore } bind def
/DiaE { stroke [] 0 setdash vpt add M
  hpt neg vpt neg V hpt vpt neg V
  hpt vpt V hpt neg vpt V closepath stroke } def
/BoxE { stroke [] 0 setdash exch hpt sub exch vpt add M
  0 vpt2 neg V hpt2 0 V 0 vpt2 V
  hpt2 neg 0 V closepath stroke } def
/TriUE { stroke [] 0 setdash vpt 1.12 mul add M
  hpt neg vpt -1.62 mul V
  hpt 2 mul 0 V
  hpt neg vpt 1.62 mul V closepath stroke } def
/TriDE { stroke [] 0 setdash vpt 1.12 mul sub M
  hpt neg vpt 1.62 mul V
  hpt 2 mul 0 V
  hpt neg vpt -1.62 mul V closepath stroke } def
/PentE { stroke [] 0 setdash gsave
  translate 0 hpt M 4 {72 rotate 0 hpt L} repeat
  closepath stroke grestore } def
/CircE { stroke [] 0 setdash 
  hpt 0 360 arc stroke } def
/Opaque { gsave closepath 1 setgray fill grestore 0 setgray closepath } def
/DiaW { stroke [] 0 setdash vpt add M
  hpt neg vpt neg V hpt vpt neg V
  hpt vpt V hpt neg vpt V Opaque stroke } def
/BoxW { stroke [] 0 setdash exch hpt sub exch vpt add M
  0 vpt2 neg V hpt2 0 V 0 vpt2 V
  hpt2 neg 0 V Opaque stroke } def
/TriUW { stroke [] 0 setdash vpt 1.12 mul add M
  hpt neg vpt -1.62 mul V
  hpt 2 mul 0 V
  hpt neg vpt 1.62 mul V Opaque stroke } def
/TriDW { stroke [] 0 setdash vpt 1.12 mul sub M
  hpt neg vpt 1.62 mul V
  hpt 2 mul 0 V
  hpt neg vpt -1.62 mul V Opaque stroke } def
/PentW { stroke [] 0 setdash gsave
  translate 0 hpt M 4 {72 rotate 0 hpt L} repeat
  Opaque stroke grestore } def
/CircW { stroke [] 0 setdash 
  hpt 0 360 arc Opaque stroke } def
/BoxFill { gsave Rec 1 setgray fill grestore } def
end
}}%
\begin{picture}(3600,2160)(0,0)%
{\GNUPLOTspecial{"
gnudict begin
gsave
0 0 translate
0.100 0.100 scale
0 setgray
newpath
1.000 UL
LTb
450 300 M
63 0 V
2937 0 R
-63 0 V
450 520 M
63 0 V
2937 0 R
-63 0 V
450 740 M
63 0 V
2937 0 R
-63 0 V
450 960 M
63 0 V
2937 0 R
-63 0 V
450 1180 M
63 0 V
2937 0 R
-63 0 V
450 1400 M
63 0 V
2937 0 R
-63 0 V
450 1620 M
63 0 V
2937 0 R
-63 0 V
450 1840 M
63 0 V
2937 0 R
-63 0 V
450 2060 M
63 0 V
2937 0 R
-63 0 V
450 300 M
0 63 V
0 1697 R
0 -63 V
825 300 M
0 63 V
0 1697 R
0 -63 V
1200 300 M
0 63 V
0 1697 R
0 -63 V
1575 300 M
0 63 V
0 1697 R
0 -63 V
1950 300 M
0 63 V
0 1697 R
0 -63 V
2325 300 M
0 63 V
0 1697 R
0 -63 V
2700 300 M
0 63 V
0 1697 R
0 -63 V
3075 300 M
0 63 V
0 1697 R
0 -63 V
3450 300 M
0 63 V
0 1697 R
0 -63 V
1.000 UL
LTb
450 300 M
3000 0 V
0 1760 V
-3000 0 V
450 300 L
0.600 UP
1.000 UL
LT0
450 1623 M
209 -21 V
520 -448 V
1440 864 L
1570 759 L
130 -34 V
260 96 V
261 129 V
521 172 V
521 94 V
187 18 V
659 1602 Pls
1179 1154 Pls
1440 864 Pls
1570 759 Pls
1700 725 Pls
1960 821 Pls
2221 950 Pls
2742 1122 Pls
3263 1216 Pls
1.000 UL
LT1
450 1475 M
209 37 V
520 50 V
261 -217 V
1700 841 L
1830 674 L
131 35 V
260 205 V
521 198 V
521 94 V
187 21 V
1.000 UL
LT2
1179 1456 M
261 52 V
260 -273 V
1830 706 L
65 -92 V
66 14 V
260 300 V
260 95 V
1.000 UL
LT3
1440 1666 M
260 -147 V
65 -495 V
65 -344 V
33 99 V
32 -233 V
66 53 V
260 183 V
0.600 UP
1.000 UL
LT4
659 1497 M
0 31 V
-31 -31 R
62 0 V
-62 31 R
62 0 V
489 21 R
0 26 V
-31 -26 R
62 0 V
-62 26 R
62 0 V
230 -233 R
0 6 V
-31 -6 R
62 0 V
-62 6 R
62 0 V
1700 838 M
0 7 V
-31 -7 R
62 0 V
-62 7 R
62 0 V
99 -174 R
0 6 V
-31 -6 R
62 0 V
-62 6 R
62 0 V
100 31 R
0 2 V
-31 -2 R
62 0 V
-62 2 R
62 0 V
229 202 R
0 4 V
-31 -4 R
62 0 V
-62 4 R
62 0 V
490 188 R
0 16 V
-31 -16 R
62 0 V
-62 16 R
62 0 V
490 83 R
0 7 V
-31 -7 R
62 0 V
-62 7 R
62 0 V
659 1512 Crs
1179 1562 Crs
1440 1345 Crs
1700 841 Crs
1830 674 Crs
1961 709 Crs
2221 914 Crs
2742 1112 Crs
3263 1206 Crs
0.600 UP
1.000 UL
LT5
1179 1436 M
0 40 V
-31 -40 R
62 0 V
-62 40 R
62 0 V
230 4 R
0 55 V
-31 -55 R
62 0 V
-62 55 R
62 0 V
229 -351 R
0 102 V
-31 -102 R
62 0 V
-62 102 R
62 0 V
99 -598 R
0 36 V
-31 -36 R
62 0 V
-62 36 R
62 0 V
34 -160 R
0 99 V
-31 -99 R
62 0 V
-62 99 R
62 0 V
35 -64 R
0 57 V
-31 -57 R
62 0 V
-62 57 R
62 0 V
229 243 R
0 58 V
-31 -58 R
62 0 V
-62 58 R
62 0 V
229 53 R
0 26 V
-31 -26 R
62 0 V
-62 26 R
62 0 V
1179 1456 Star
1440 1508 Star
1700 1235 Star
1830 706 Star
1895 614 Star
1961 628 Star
2221 928 Star
2481 1023 Star
0.600 UP
1.000 UL
LT6
1440 1490 M
0 352 V
-31 -352 R
62 0 V
-62 352 R
62 0 V
229 -475 R
0 304 V
-31 -304 R
62 0 V
-62 304 R
62 0 V
34 -741 R
0 187 V
1734 930 M
62 0 V
-62 187 R
62 0 V
34 -596 R
0 317 V
1799 521 M
62 0 V
-62 317 R
62 0 V
2 -149 R
0 179 V
1832 689 M
62 0 V
-62 179 R
62 0 V
1 -454 R
0 264 V
1864 414 M
62 0 V
-62 264 R
62 0 V
35 -201 R
0 244 V
1930 477 M
62 0 V
-62 244 R
62 0 V
229 -32 R
0 185 V
2190 689 M
62 0 V
-62 185 R
62 0 V
1440 1666 Box
1700 1519 Box
1765 1024 Box
1830 680 Box
1863 779 Box
1895 546 Box
1961 599 Box
2221 782 Box
stroke
grestore
end
showpage
}}%
\put(1950,50){\makebox(0,0){$\gamma$}}%
\put(100,1180){%
\makebox(0,0)[b]{\shortstack{$P_3$}}%
}%
\put(3450,200){\makebox(0,0){0.65}}%
\put(3075,200){\makebox(0,0){0.6}}%
\put(2700,200){\makebox(0,0){0.55}}%
\put(2325,200){\makebox(0,0){0.5}}%
\put(1950,200){\makebox(0,0){0.45}}%
\put(1575,200){\makebox(0,0){0.4}}%
\put(1200,200){\makebox(0,0){0.35}}%
\put(825,200){\makebox(0,0){0.3}}%
\put(450,200){\makebox(0,0){0.25}}%
\put(400,2060){\makebox(0,0)[r]{0.6}}%
\put(400,1840){\makebox(0,0)[r]{0.4}}%
\put(400,1620){\makebox(0,0)[r]{0.2}}%
\put(400,1400){\makebox(0,0)[r]{0}}%
\put(400,1180){\makebox(0,0)[r]{-0.2}}%
\put(400,960){\makebox(0,0)[r]{-0.4}}%
\put(400,740){\makebox(0,0)[r]{-0.6}}%
\put(400,520){\makebox(0,0)[r]{-0.8}}%
\put(400,300){\makebox(0,0)[r]{-1}}%
\end{picture}%
\endgroup
 
}
\end	{center}
\caption{The cubic local plaquette correlator, $P_3$, as a 
function of $\gamma=\frac{\beta}{2N^2}$
for SU(6) ($+$), SU(12) ($\times$), 
SU(24) ($\ast$) and SU(48) ($\Box$).}  
\label{fig_P_3}
\end 	{figure}

To investigate this further, we show in Fig.~\ref{fig_P_3} our 
results for $P_3$ (the `local' version of $C_3$) for $SU(6)$, 
$SU(12)$, $SU(24)$ and $SU(48)$. There is clearly an increasingly
sharp transition as $N$ increases around $\gamma\simeq 0.43$. This behaviour is remarkably similar to what happens in $D=1+1$.

Finally, if we compare plaquette eigenvalue densities across the $D=2+1$ and $D=1+1$ transitions directly,
we find that they are very similar both below and above the
transition.

Despite these striking similarities, when we look in more
detail we observe significant differences 
between the bulk transition in 2+1 dimensions and the 
Gross--Witten transition. In particular there is
a peak in the specific heat in $D=2+1$ which is
not present in $D=1+1$. To investigate this we calculate the contribution to the specific heat $C_2$ from correlations between a plaquette and its 
neighbours which share an edge but are not in the same plane,  
$C_o$. We find a clear 
peak, growing with $N$, in our results, plotted 
in Fig.~\ref{fig_outplane}. We see a similar peak, but almost exactly a factor of four lower, in $C_f$, the contribution from correlations 
between a plaquette and the plaquettes facing it across an 
elementary cube, as expected
if the correlations are due to a flux emerging from the cube
symmetrically through every face, i.e. due to the presence
of monopole--instantons.

There are several scenarios for what happens at $N=\infty$ that are consistent with our results. One possibility is a third--order phase transition with critical exponent $\alpha$ different from $-1$. Alternatively there could be a second--order phase transition  driven either by local fluctuations or by the correlation length diverging (or both). Our results cannot distinguish between these scenarios.

To search for the possibility of a diverging correlation length, we measured the mass of the lightest particle that couples to the plaquette, in both $SU(6)$ and $SU(12)$. Our results show a modest dip in the  
masses  near the transition, which becomes more significant 
as we increase $N$.
However, the masses are large, and if the correlation length
is going to show any sign of diverging it 
will be at much larger values of $N$ than are accessible
to our calculations.

\subsection{Wilson loops}
When we calculate how $\langle u_w \rangle$ varies with $\lambda$ 
we see no sign of any singularity developing in this quantity, 
or in our simulataneous calculations of 
$\partial \langle u_w \rangle / \partial\lambda$.
The  more accurately calculated local version  
of the correlator that is equivalent to the derivative, also
shows no evidence of developing the sort of 
cusp that might suggest an $N=\infty$ singularity
in the second derivative.
We have also looked at quantities analogous to
$P_2$ and $P_3$ for the plaquette. The variation of these
quantities with $\gamma$ also does not become sharper with $N$.
All our results are in fact essentially identical to those 
we obtain in similar calculations in $D=1+1$.

It is possible
that there are more subtle non-analyticities associated with a gap forming in the
eigenvalue spectrum, of the kind that
exist in $D=1+1$.
To search for such behaviour we directly compare Wilson loop eigenvalue spectra
in 1+1 and 2+1 dimensions. We first evaluate the 
spectrum in 1+1 dimensions at the critical coupling at which 
the gap forms. A true gap only forms at $N=\infty$; for finite
$N$ we use the same value of the 't~Hooft coupling,
$\lambda_c = \frac{1}{\gamma_c} = 4(1-e^{\frac{-2a^2}{A}})$,
where $A$ is the area of the Wilson loop in physical units.
Having obtained the spectrum (numerically) in 
$D=1+1$ we then calculate the 
spectrum in $D=2+1$ for the same size loop and for the same $N$,
varying the coupling to a value where the two spectra match.

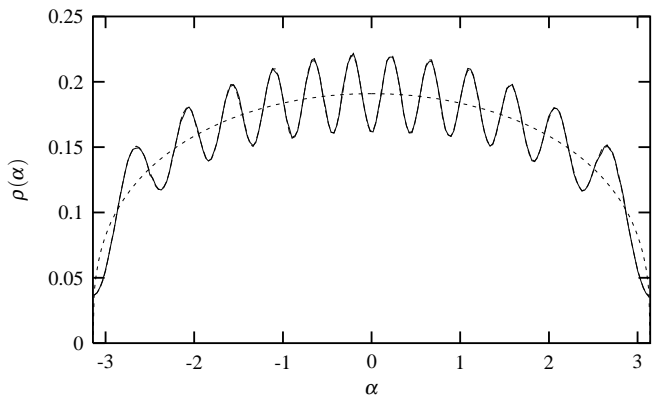
\begin	{figure}[p]
\begin	{center}
\leavevmode
\scalebox{0.7}{
\begingroup%
  \makeatletter%
  \newcommand{\GNUPLOTspecial}{%
    \@sanitize\catcode`\%=14\relax\special}%
  \setlength{\unitlength}{0.1bp}%
{\GNUPLOTspecial{!
/gnudict 256 dict def
gnudict begin
/Color false def
/Solid false def
/gnulinewidth 5.000 def
/userlinewidth gnulinewidth def
/vshift -33 def
/dl {10 mul} def
/hpt_ 31.5 def
/vpt_ 31.5 def
/hpt hpt_ def
/vpt vpt_ def
/M {moveto} bind def
/L {lineto} bind def
/R {rmoveto} bind def
/V {rlineto} bind def
/vpt2 vpt 2 mul def
/hpt2 hpt 2 mul def
/Lshow { currentpoint stroke M
  0 vshift R show } def
/Rshow { currentpoint stroke M
  dup stringwidth pop neg vshift R show } def
/Cshow { currentpoint stroke M
  dup stringwidth pop -2 div vshift R show } def
/UP { dup vpt_ mul /vpt exch def hpt_ mul /hpt exch def
  /hpt2 hpt 2 mul def /vpt2 vpt 2 mul def } def
/DL { Color {setrgbcolor Solid {pop []} if 0 setdash }
 {pop pop pop Solid {pop []} if 0 setdash} ifelse } def
/BL { stroke userlinewidth 2 mul setlinewidth } def
/AL { stroke userlinewidth 2 div setlinewidth } def
/UL { dup gnulinewidth mul /userlinewidth exch def
      10 mul /udl exch def } def
/PL { stroke userlinewidth setlinewidth } def
/LTb { BL [] 0 0 0 DL } def
/LTa { AL [1 udl mul 2 udl mul] 0 setdash 0 0 0 setrgbcolor } def
/LT0 { PL [] 1 0 0 DL } def
/LT1 { PL [4 dl 2 dl] 0 1 0 DL } def
/LT2 { PL [2 dl 3 dl] 0 0 1 DL } def
/LT3 { PL [1 dl 1.5 dl] 1 0 1 DL } def
/LT4 { PL [5 dl 2 dl 1 dl 2 dl] 0 1 1 DL } def
/LT5 { PL [4 dl 3 dl 1 dl 3 dl] 1 1 0 DL } def
/LT6 { PL [2 dl 2 dl 2 dl 4 dl] 0 0 0 DL } def
/LT7 { PL [2 dl 2 dl 2 dl 2 dl 2 dl 4 dl] 1 0.3 0 DL } def
/LT8 { PL [2 dl 2 dl 2 dl 2 dl 2 dl 2 dl 2 dl 4 dl] 0.5 0.5 0.5 DL } def
/Pnt { stroke [] 0 setdash
   gsave 1 setlinecap M 0 0 V stroke grestore } def
/Dia { stroke [] 0 setdash 2 copy vpt add M
  hpt neg vpt neg V hpt vpt neg V
  hpt vpt V hpt neg vpt V closepath stroke
  Pnt } def
/Pls { stroke [] 0 setdash vpt sub M 0 vpt2 V
  currentpoint stroke M
  hpt neg vpt neg R hpt2 0 V stroke
  } def
/Box { stroke [] 0 setdash 2 copy exch hpt sub exch vpt add M
  0 vpt2 neg V hpt2 0 V 0 vpt2 V
  hpt2 neg 0 V closepath stroke
  Pnt } def
/Crs { stroke [] 0 setdash exch hpt sub exch vpt add M
  hpt2 vpt2 neg V currentpoint stroke M
  hpt2 neg 0 R hpt2 vpt2 V stroke } def
/TriU { stroke [] 0 setdash 2 copy vpt 1.12 mul add M
  hpt neg vpt -1.62 mul V
  hpt 2 mul 0 V
  hpt neg vpt 1.62 mul V closepath stroke
  Pnt  } def
/Star { 2 copy Pls Crs } def
/BoxF { stroke [] 0 setdash exch hpt sub exch vpt add M
  0 vpt2 neg V  hpt2 0 V  0 vpt2 V
  hpt2 neg 0 V  closepath fill } def
/TriUF { stroke [] 0 setdash vpt 1.12 mul add M
  hpt neg vpt -1.62 mul V
  hpt 2 mul 0 V
  hpt neg vpt 1.62 mul V closepath fill } def
/TriD { stroke [] 0 setdash 2 copy vpt 1.12 mul sub M
  hpt neg vpt 1.62 mul V
  hpt 2 mul 0 V
  hpt neg vpt -1.62 mul V closepath stroke
  Pnt  } def
/TriDF { stroke [] 0 setdash vpt 1.12 mul sub M
  hpt neg vpt 1.62 mul V
  hpt 2 mul 0 V
  hpt neg vpt -1.62 mul V closepath fill} def
/DiaF { stroke [] 0 setdash vpt add M
  hpt neg vpt neg V hpt vpt neg V
  hpt vpt V hpt neg vpt V closepath fill } def
/Pent { stroke [] 0 setdash 2 copy gsave
  translate 0 hpt M 4 {72 rotate 0 hpt L} repeat
  closepath stroke grestore Pnt } def
/PentF { stroke [] 0 setdash gsave
  translate 0 hpt M 4 {72 rotate 0 hpt L} repeat
  closepath fill grestore } def
/Circle { stroke [] 0 setdash 2 copy
  hpt 0 360 arc stroke Pnt } def
/CircleF { stroke [] 0 setdash hpt 0 360 arc fill } def
/C0 { BL [] 0 setdash 2 copy moveto vpt 90 450  arc } bind def
/C1 { BL [] 0 setdash 2 copy        moveto
       2 copy  vpt 0 90 arc closepath fill
               vpt 0 360 arc closepath } bind def
/C2 { BL [] 0 setdash 2 copy moveto
       2 copy  vpt 90 180 arc closepath fill
               vpt 0 360 arc closepath } bind def
/C3 { BL [] 0 setdash 2 copy moveto
       2 copy  vpt 0 180 arc closepath fill
               vpt 0 360 arc closepath } bind def
/C4 { BL [] 0 setdash 2 copy moveto
       2 copy  vpt 180 270 arc closepath fill
               vpt 0 360 arc closepath } bind def
/C5 { BL [] 0 setdash 2 copy moveto
       2 copy  vpt 0 90 arc
       2 copy moveto
       2 copy  vpt 180 270 arc closepath fill
               vpt 0 360 arc } bind def
/C6 { BL [] 0 setdash 2 copy moveto
      2 copy  vpt 90 270 arc closepath fill
              vpt 0 360 arc closepath } bind def
/C7 { BL [] 0 setdash 2 copy moveto
      2 copy  vpt 0 270 arc closepath fill
              vpt 0 360 arc closepath } bind def
/C8 { BL [] 0 setdash 2 copy moveto
      2 copy vpt 270 360 arc closepath fill
              vpt 0 360 arc closepath } bind def
/C9 { BL [] 0 setdash 2 copy moveto
      2 copy  vpt 270 450 arc closepath fill
              vpt 0 360 arc closepath } bind def
/C10 { BL [] 0 setdash 2 copy 2 copy moveto vpt 270 360 arc closepath fill
       2 copy moveto
       2 copy vpt 90 180 arc closepath fill
               vpt 0 360 arc closepath } bind def
/C11 { BL [] 0 setdash 2 copy moveto
       2 copy  vpt 0 180 arc closepath fill
       2 copy moveto
       2 copy  vpt 270 360 arc closepath fill
               vpt 0 360 arc closepath } bind def
/C12 { BL [] 0 setdash 2 copy moveto
       2 copy  vpt 180 360 arc closepath fill
               vpt 0 360 arc closepath } bind def
/C13 { BL [] 0 setdash  2 copy moveto
       2 copy  vpt 0 90 arc closepath fill
       2 copy moveto
       2 copy  vpt 180 360 arc closepath fill
               vpt 0 360 arc closepath } bind def
/C14 { BL [] 0 setdash 2 copy moveto
       2 copy  vpt 90 360 arc closepath fill
               vpt 0 360 arc } bind def
/C15 { BL [] 0 setdash 2 copy vpt 0 360 arc closepath fill
               vpt 0 360 arc closepath } bind def
/Rec   { newpath 4 2 roll moveto 1 index 0 rlineto 0 exch rlineto
       neg 0 rlineto closepath } bind def
/Square { dup Rec } bind def
/Bsquare { vpt sub exch vpt sub exch vpt2 Square } bind def
/S0 { BL [] 0 setdash 2 copy moveto 0 vpt rlineto BL Bsquare } bind def
/S1 { BL [] 0 setdash 2 copy vpt Square fill Bsquare } bind def
/S2 { BL [] 0 setdash 2 copy exch vpt sub exch vpt Square fill Bsquare } bind def
/S3 { BL [] 0 setdash 2 copy exch vpt sub exch vpt2 vpt Rec fill Bsquare } bind def
/S4 { BL [] 0 setdash 2 copy exch vpt sub exch vpt sub vpt Square fill Bsquare } bind def
/S5 { BL [] 0 setdash 2 copy 2 copy vpt Square fill
       exch vpt sub exch vpt sub vpt Square fill Bsquare } bind def
/S6 { BL [] 0 setdash 2 copy exch vpt sub exch vpt sub vpt vpt2 Rec fill Bsquare } bind def
/S7 { BL [] 0 setdash 2 copy exch vpt sub exch vpt sub vpt vpt2 Rec fill
       2 copy vpt Square fill
       Bsquare } bind def
/S8 { BL [] 0 setdash 2 copy vpt sub vpt Square fill Bsquare } bind def
/S9 { BL [] 0 setdash 2 copy vpt sub vpt vpt2 Rec fill Bsquare } bind def
/S10 { BL [] 0 setdash 2 copy vpt sub vpt Square fill 2 copy exch vpt sub exch vpt Square fill
       Bsquare } bind def
/S11 { BL [] 0 setdash 2 copy vpt sub vpt Square fill 2 copy exch vpt sub exch vpt2 vpt Rec fill
       Bsquare } bind def
/S12 { BL [] 0 setdash 2 copy exch vpt sub exch vpt sub vpt2 vpt Rec fill Bsquare } bind def
/S13 { BL [] 0 setdash 2 copy exch vpt sub exch vpt sub vpt2 vpt Rec fill
       2 copy vpt Square fill Bsquare } bind def
/S14 { BL [] 0 setdash 2 copy exch vpt sub exch vpt sub vpt2 vpt Rec fill
       2 copy exch vpt sub exch vpt Square fill Bsquare } bind def
/S15 { BL [] 0 setdash 2 copy Bsquare fill Bsquare } bind def
/D0 { gsave translate 45 rotate 0 0 S0 stroke grestore } bind def
/D1 { gsave translate 45 rotate 0 0 S1 stroke grestore } bind def
/D2 { gsave translate 45 rotate 0 0 S2 stroke grestore } bind def
/D3 { gsave translate 45 rotate 0 0 S3 stroke grestore } bind def
/D4 { gsave translate 45 rotate 0 0 S4 stroke grestore } bind def
/D5 { gsave translate 45 rotate 0 0 S5 stroke grestore } bind def
/D6 { gsave translate 45 rotate 0 0 S6 stroke grestore } bind def
/D7 { gsave translate 45 rotate 0 0 S7 stroke grestore } bind def
/D8 { gsave translate 45 rotate 0 0 S8 stroke grestore } bind def
/D9 { gsave translate 45 rotate 0 0 S9 stroke grestore } bind def
/D10 { gsave translate 45 rotate 0 0 S10 stroke grestore } bind def
/D11 { gsave translate 45 rotate 0 0 S11 stroke grestore } bind def
/D12 { gsave translate 45 rotate 0 0 S12 stroke grestore } bind def
/D13 { gsave translate 45 rotate 0 0 S13 stroke grestore } bind def
/D14 { gsave translate 45 rotate 0 0 S14 stroke grestore } bind def
/D15 { gsave translate 45 rotate 0 0 S15 stroke grestore } bind def
/DiaE { stroke [] 0 setdash vpt add M
  hpt neg vpt neg V hpt vpt neg V
  hpt vpt V hpt neg vpt V closepath stroke } def
/BoxE { stroke [] 0 setdash exch hpt sub exch vpt add M
  0 vpt2 neg V hpt2 0 V 0 vpt2 V
  hpt2 neg 0 V closepath stroke } def
/TriUE { stroke [] 0 setdash vpt 1.12 mul add M
  hpt neg vpt -1.62 mul V
  hpt 2 mul 0 V
  hpt neg vpt 1.62 mul V closepath stroke } def
/TriDE { stroke [] 0 setdash vpt 1.12 mul sub M
  hpt neg vpt 1.62 mul V
  hpt 2 mul 0 V
  hpt neg vpt -1.62 mul V closepath stroke } def
/PentE { stroke [] 0 setdash gsave
  translate 0 hpt M 4 {72 rotate 0 hpt L} repeat
  closepath stroke grestore } def
/CircE { stroke [] 0 setdash 
  hpt 0 360 arc stroke } def
/Opaque { gsave closepath 1 setgray fill grestore 0 setgray closepath } def
/DiaW { stroke [] 0 setdash vpt add M
  hpt neg vpt neg V hpt vpt neg V
  hpt vpt V hpt neg vpt V Opaque stroke } def
/BoxW { stroke [] 0 setdash exch hpt sub exch vpt add M
  0 vpt2 neg V hpt2 0 V 0 vpt2 V
  hpt2 neg 0 V Opaque stroke } def
/TriUW { stroke [] 0 setdash vpt 1.12 mul add M
  hpt neg vpt -1.62 mul V
  hpt 2 mul 0 V
  hpt neg vpt 1.62 mul V Opaque stroke } def
/TriDW { stroke [] 0 setdash vpt 1.12 mul sub M
  hpt neg vpt 1.62 mul V
  hpt 2 mul 0 V
  hpt neg vpt -1.62 mul V Opaque stroke } def
/PentW { stroke [] 0 setdash gsave
  translate 0 hpt M 4 {72 rotate 0 hpt L} repeat
  Opaque stroke grestore } def
/CircW { stroke [] 0 setdash 
  hpt 0 360 arc Opaque stroke } def
/BoxFill { gsave Rec 1 setgray fill grestore } def
end
}}%
\begin{picture}(3600,2160)(0,0)%
{\GNUPLOTspecial{"
gnudict begin
gsave
0 0 translate
0.100 0.100 scale
0 setgray
newpath
1.000 UL
LTb
450 300 M
63 0 V
2937 0 R
-63 0 V
450 652 M
63 0 V
2937 0 R
-63 0 V
450 1004 M
63 0 V
2937 0 R
-63 0 V
450 1356 M
63 0 V
2937 0 R
-63 0 V
450 1708 M
63 0 V
2937 0 R
-63 0 V
450 2060 M
63 0 V
2937 0 R
-63 0 V
518 300 M
0 63 V
0 1697 R
0 -63 V
995 300 M
0 63 V
0 1697 R
0 -63 V
1473 300 M
0 63 V
0 1697 R
0 -63 V
1950 300 M
0 63 V
0 1697 R
0 -63 V
2427 300 M
0 63 V
0 1697 R
0 -63 V
2905 300 M
0 63 V
0 1697 R
0 -63 V
3382 300 M
0 63 V
0 1697 R
0 -63 V
1.000 UL
LTb
450 300 M
3000 0 V
0 1760 V
-3000 0 V
450 300 L
1.000 UL
LT0
457 559 M
15 12 V
15 28 V
15 39 V
15 54 V
15 66 V
15 74 V
15 69 V
15 85 V
15 83 V
15 73 V
15 74 V
15 64 V
15 39 V
15 26 V
15 5 V
15 2 V
15 -17 V
15 -27 V
15 -42 V
15 -61 V
15 -16 V
15 -37 V
15 -21 V
15 -6 V
15 14 V
15 31 V
15 53 V
15 76 V
15 55 V
15 73 V
15 64 V
15 32 V
15 37 V
15 12 V
15 -21 V
15 -47 V
15 -46 V
15 -60 V
15 -57 V
15 -44 V
15 -16 V
15 12 V
15 40 V
15 46 V
15 67 V
15 83 V
15 76 V
15 44 V
15 44 V
15 -1 V
15 -24 V
15 -43 V
15 -76 V
15 -78 V
15 -60 V
15 -31 V
15 -16 V
15 17 V
15 58 V
15 75 V
15 81 V
15 83 V
15 63 V
15 38 V
15 -18 V
15 -17 V
15 -66 V
15 -86 V
15 -86 V
15 -46 V
15 -56 V
15 8 V
15 32 V
15 78 V
15 89 V
15 92 V
15 62 V
15 59 V
15 9 V
15 -42 V
15 -59 V
15 -95 V
15 -67 V
15 -85 V
15 -49 V
15 -4 V
15 38 V
15 65 V
15 119 V
15 77 V
15 75 V
15 32 V
15 23 V
15 -43 V
15 -81 V
15 -92 V
15 -86 V
15 -64 V
15 -54 V
16 -3 V
15 48 V
15 84 V
15 72 V
15 105 V
15 75 V
15 19 V
15 0 V
15 -44 V
15 -69 V
15 -83 V
15 -107 V
15 -67 V
15 -39 V
15 2 V
15 56 V
15 66 V
15 92 V
15 83 V
15 60 V
15 24 V
15 5 V
15 -45 V
15 -73 V
15 -82 V
15 -95 V
15 -75 V
15 -28 V
15 -8 V
15 30 V
15 67 V
15 78 V
15 76 V
15 67 V
15 48 V
15 -4 V
15 -31 V
15 -71 V
15 -65 V
15 -104 V
15 -67 V
15 -51 V
15 -26 V
15 14 V
15 46 V
15 62 V
15 62 V
15 70 V
15 59 V
15 10 V
15 10 V
15 -41 V
15 -58 V
15 -61 V
15 -84 V
15 -77 V
15 -44 V
15 -46 V
15 -5 V
15 25 V
15 39 V
15 56 V
15 61 V
15 61 V
15 28 V
15 17 V
15 -2 V
15 -20 V
15 -49 V
15 -68 V
15 -73 V
15 -64 V
15 -72 V
15 -45 V
15 -29 V
15 -24 V
15 6 V
15 20 V
15 35 V
15 47 V
15 43 V
15 29 V
15 24 V
15 34 V
15 11 V
15 -25 V
15 -17 V
15 -47 V
15 -77 V
15 -51 V
15 -70 V
15 -100 V
15 -80 V
15 -73 V
15 -71 V
15 -69 V
15 -47 V
15 -44 V
15 -29 V
15 -19 V
1.000 UL
LT1
457 558 M
15 13 V
16 25 V
15 41 V
15 53 V
14 67 V
15 77 V
16 76 V
15 85 V
14 77 V
15 74 V
16 74 V
14 58 V
16 42 V
14 30 V
16 11 V
15 -7 V
14 -16 V
16 -33 V
14 -42 V
15 -44 V
16 -37 V
14 -35 V
15 -19 V
16 0 V
14 15 V
15 29 V
16 55 V
14 63 V
15 72 V
16 64 V
14 73 V
15 41 V
16 26 V
14 4 V
15 -21 V
16 -51 V
14 -52 V
15 -50 V
16 -55 V
15 -41 V
14 -18 V
16 10 V
15 38 V
14 49 V
16 81 V
15 72 V
14 70 V
16 60 V
15 23 V
14 6 V
15 -30 V
15 -56 V
16 -64 V
15 -64 V
15 -63 V
14 -44 V
15 -10 V
15 27 V
16 58 V
15 72 V
15 76 V
14 85 V
15 57 V
15 43 V
15 -1 V
16 -40 V
15 -67 V
15 -76 V
14 -84 V
15 -71 V
15 -29 V
16 7 V
15 33 V
15 71 V
14 94 V
15 88 V
15 74 V
16 42 V
15 3 V
15 -30 V
14 -61 V
15 -86 V
15 -91 V
16 -73 V
15 -44 V
15 -5 V
14 38 V
15 75 V
15 86 V
15 95 V
16 84 V
15 39 V
15 -2 V
14 -38 V
15 -68 V
15 -91 V
16 -92 V
15 -79 V
15 -34 V
14 -4 V
15 45 V
15 69 V
15 100 V
15 82 V
15 82 V
16 26 V
15 3 V
15 -47 V
15 -78 V
15 -90 V
15 -96 V
15 -67 V
14 -35 V
15 4 V
15 41 V
15 75 V
15 81 V
15 94 V
16 66 V
15 29 V
15 1 V
15 -56 V
15 -70 V
15 -91 V
14 -90 V
15 -66 V
15 -31 V
15 -8 V
15 32 V
15 63 V
16 84 V
15 71 V
15 68 V
15 37 V
15 6 V
15 -37 V
15 -63 V
14 -74 V
15 -85 V
15 -78 V
15 -45 V
15 -28 V
15 4 V
16 46 V
15 62 V
15 69 V
15 66 V
15 56 V
15 23 V
14 1 V
15 -39 V
15 -50 V
15 -73 V
15 -80 V
15 -67 V
16 -63 V
15 -31 V
15 -9 V
15 19 V
15 43 V
15 55 V
15 61 V
14 47 V
15 39 V
15 25 V
15 -6 V
15 -22 V
15 -46 V
16 -65 V
15 -69 V
15 -68 V
15 -65 V
15 -51 V
15 -36 V
14 -20 V
15 5 V
15 20 V
15 30 V
15 43 V
15 51 V
16 33 V
15 38 V
15 12 V
15 7 V
15 -10 V
15 -30 V
15 -45 V
14 -56 V
15 -65 V
15 -82 V
15 -85 V
15 -78 V
15 -78 V
16 -75 V
15 -65 V
15 -48 V
15 -44 V
15 -29 V
15 -15 V
1.000 UL
LT2
1950 1644 M
8 0 V
7 0 V
8 0 V
7 0 V
8 0 V
7 0 V
8 0 V
7 -1 V
8 0 V
7 0 V
8 0 V
7 -1 V
8 0 V
7 0 V
8 -1 V
7 0 V
8 0 V
7 -1 V
8 0 V
7 -1 V
8 0 V
7 -1 V
8 0 V
7 -1 V
8 0 V
7 -1 V
8 -1 V
7 0 V
8 -1 V
7 -1 V
8 -1 V
7 0 V
8 -1 V
7 -1 V
8 -1 V
7 -1 V
8 -1 V
7 -1 V
8 -1 V
7 -1 V
8 -1 V
7 -1 V
8 -1 V
7 -1 V
8 -1 V
7 -1 V
8 -1 V
7 -1 V
8 -2 V
7 -1 V
8 -1 V
7 -2 V
8 -1 V
7 -1 V
8 -2 V
7 -1 V
8 -1 V
7 -2 V
8 -1 V
7 -2 V
8 -2 V
7 -1 V
8 -2 V
7 -1 V
8 -2 V
7 -2 V
8 -2 V
7 -1 V
8 -2 V
7 -2 V
8 -2 V
7 -2 V
8 -2 V
7 -2 V
8 -2 V
7 -2 V
8 -2 V
7 -2 V
8 -2 V
7 -2 V
8 -2 V
7 -2 V
8 -3 V
7 -2 V
8 -2 V
7 -3 V
8 -2 V
7 -2 V
8 -3 V
7 -2 V
8 -3 V
7 -2 V
8 -3 V
7 -3 V
8 -2 V
7 -3 V
8 -3 V
7 -3 V
8 -2 V
7 -3 V
8 -3 V
7 -3 V
8 -3 V
7 -3 V
8 -3 V
7 -3 V
8 -4 V
7 -3 V
8 -3 V
7 -3 V
8 -4 V
7 -3 V
8 -4 V
7 -3 V
8 -4 V
7 -3 V
8 -4 V
7 -4 V
8 -3 V
7 -4 V
8 -4 V
7 -4 V
8 -4 V
7 -4 V
8 -4 V
7 -4 V
8 -4 V
7 -4 V
8 -5 V
7 -4 V
8 -5 V
7 -4 V
8 -5 V
7 -4 V
8 -5 V
7 -5 V
8 -5 V
7 -4 V
7 -5 V
8 -5 V
8 -6 V
7 -5 V
8 -5 V
7 -6 V
8 -5 V
7 -6 V
8 -5 V
7 -6 V
8 -6 V
7 -6 V
8 -6 V
7 -6 V
8 -6 V
7 -7 V
8 -6 V
7 -7 V
8 -7 V
7 -7 V
8 -7 V
7 -7 V
8 -7 V
7 -8 V
8 -7 V
7 -8 V
8 -8 V
7 -8 V
8 -8 V
7 -9 V
8 -9 V
7 -9 V
8 -9 V
7 -9 V
8 -10 V
7 -10 V
8 -10 V
7 -10 V
8 -11 V
7 -11 V
8 -12 V
7 -12 V
8 -12 V
7 -13 V
8 -13 V
7 -14 V
8 -15 V
7 -15 V
8 -16 V
7 -17 V
8 -18 V
7 -19 V
8 -21 V
7 -22 V
8 -24 V
7 -26 V
8 -30 V
7 -34 V
8 -40 V
7 -51 V
8 -72 V
7 -278 V
1.000 UL
LT2
1950 1644 M
-7 0 V
-8 0 V
-7 0 V
-8 0 V
-7 0 V
-8 0 V
-7 0 V
-8 -1 V
-7 0 V
-8 0 V
-7 0 V
-8 -1 V
-7 0 V
-8 0 V
-7 -1 V
-8 0 V
-7 0 V
-8 -1 V
-7 0 V
-8 -1 V
-7 0 V
-8 -1 V
-7 0 V
-8 -1 V
-7 0 V
-8 -1 V
-7 -1 V
-8 0 V
-7 -1 V
-8 -1 V
-7 -1 V
-8 0 V
-7 -1 V
-8 -1 V
-7 -1 V
-8 -1 V
-7 -1 V
-8 -1 V
-7 -1 V
-8 -1 V
-7 -1 V
-8 -1 V
-7 -1 V
-8 -1 V
-7 -1 V
-8 -1 V
-7 -1 V
-8 -1 V
-7 -2 V
-8 -1 V
-7 -1 V
-8 -2 V
-7 -1 V
-8 -1 V
-7 -2 V
-8 -1 V
-7 -1 V
-8 -2 V
-7 -1 V
-8 -2 V
-7 -2 V
-8 -1 V
-7 -2 V
-8 -1 V
-7 -2 V
-8 -2 V
-7 -2 V
-8 -1 V
-7 -2 V
-8 -2 V
-7 -2 V
-8 -2 V
-7 -2 V
-8 -2 V
-7 -2 V
-8 -2 V
-7 -2 V
-8 -2 V
-7 -2 V
-8 -2 V
-7 -2 V
-8 -2 V
-7 -3 V
-8 -2 V
-7 -2 V
-8 -3 V
-7 -2 V
-8 -2 V
-7 -3 V
-8 -2 V
-7 -3 V
-8 -2 V
-7 -3 V
-8 -3 V
-7 -2 V
-8 -3 V
-7 -3 V
-8 -3 V
-7 -2 V
-8 -3 V
-7 -3 V
-8 -3 V
-7 -3 V
-8 -3 V
-7 -3 V
-8 -3 V
-7 -4 V
-8 -3 V
-7 -3 V
-8 -3 V
-7 -4 V
-8 -3 V
-7 -4 V
-8 -3 V
-7 -4 V
-8 -3 V
-7 -4 V
-8 -4 V
-7 -3 V
-8 -4 V
-7 -4 V
-8 -4 V
-7 -4 V
-8 -4 V
-7 -4 V
-8 -4 V
-7 -4 V
-8 -4 V
-7 -5 V
-8 -4 V
-7 -5 V
-8 -4 V
-7 -5 V
-8 -4 V
-7 -5 V
-8 -5 V
-8 -5 V
-7 -4 V
-7 -5 V
-8 -5 V
-7 -6 V
-8 -5 V
-7 -5 V
-8 -6 V
-7 -5 V
-8 -6 V
-7 -5 V
-8 -6 V
-7 -6 V
-8 -6 V
-7 -6 V
-8 -6 V
-7 -6 V
-8 -7 V
-7 -6 V
-8 -7 V
-8 -7 V
-7 -7 V
-7 -7 V
-8 -7 V
-8 -7 V
-7 -8 V
-7 -7 V
-8 -8 V
-7 -8 V
-8 -8 V
-7 -8 V
-8 -9 V
-7 -9 V
-8 -9 V
-7 -9 V
-8 -9 V
-8 -10 V
-7 -10 V
-7 -10 V
-8 -10 V
-7 -11 V
-8 -11 V
-7 -12 V
-8 -12 V
-7 -12 V
-8 -13 V
-8 -13 V
-7 -14 V
-7 -15 V
-8 -15 V
-8 -16 V
-7 -17 V
-7 -18 V
-8 -19 V
-7 -21 V
-8 -22 V
-7 -24 V
-8 -26 V
-7 -30 V
-8 -34 V
-7 -40 V
-8 -51 V
-8 -72 V
450 300 L
stroke
grestore
end
showpage
}}%
\put(1950,50){\makebox(0,0){$\alpha$}}%
\put(100,1180){%
\makebox(0,0)[b]{\shortstack{$\rho(\alpha)$}}%
}%
\put(3382,200){\makebox(0,0){3}}%
\put(2905,200){\makebox(0,0){2}}%
\put(2427,200){\makebox(0,0){1}}%
\put(1950,200){\makebox(0,0){0}}%
\put(1473,200){\makebox(0,0){-1}}%
\put(995,200){\makebox(0,0){-2}}%
\put(518,200){\makebox(0,0){-3}}%
\put(400,2060){\makebox(0,0)[r]{0.25}}%
\put(400,1708){\makebox(0,0)[r]{0.2}}%
\put(400,1356){\makebox(0,0)[r]{0.15}}%
\put(400,1004){\makebox(0,0)[r]{0.1}}%
\put(400,652){\makebox(0,0)[r]{0.05}}%
\put(400,300){\makebox(0,0)[r]{0}}%
\end{picture}%
\endgroup
 
}
\end	{center}
\caption{$3\times 3$ Wilson loop eigenvalue density, $e^{i\alpha}$, 
for SU(12) in 1+1 dimensions at $\gamma=\frac{\beta}{2N^2}=1.255$ 
(solid line) and in 2+1 dimensions at $\gamma=0.722$ (long dashes),
and the continuum large--N distribution in 1+1 dimensions at
$A=A_{crit}$ (short dashes).}
\label{fig_wilsoneigenvaluedensity}
\end 	{figure}

We find that it is always possible to achieve such a match,
for any $N$ and for any size of Wilson loop.
We show an example in Fig.~\ref{fig_wilsoneigenvaluedensity},
where we compare the eigenvalue densities of the $3\times 3$ 
Wilson loops in SU(12) in 1+1 and 2+1 
dimensions. The spectra are clearly very similar and indeed indistinguishable
on this plot. We also find that the spectra can be matched 
when they are away from the critical coupling.

The fact that at finite but large $N$ we can match so precisely 
the $D=1+1$ and $D=2+1$ eigenvalue spectra provides convincing evidence that 
the Wilson loops in the $D=2+1$ $N=\infty$ theory also undergo
a transition involving the formation of a gap in the eigenvalue
spectrum. 

All the above is an immediate corollary
of a much stronger 
and rather surprising result concerning the matching of Wilson loop
eigenvalue spectra in 1+1 and 2+1 (and indeed 3+1) dimensions.

The general statement is that if we take an $n\times n$ Wilson
loop  $U_w^{n\times n}$ in the SU($N$) gauge theory and calculate
the eigenvalue spectra in $D$ and $D^\prime$ dimensions, we
find that the spectra match at the couplings $\lambda_D$ 
and $\lambda_{D^\prime}$ at which the averages of the traces 
$u_w^{n\times n}=\frac{1}{N}\mathrm{Re}\mathrm{Tr}\{U_w^{n\times n}\}$
are equal.
We have tested this matching for $D=1+1$ and $D=2+1$ over groups 
in the range $N=2$ to $N=48$ and for Wilson loops ranging in size 
from $1\times 1$ (the plaquette) to $8\times 8$.
Some sample calculations in
$D=3+1$~\cite{BurTepVai}
strongly suggest that the same is true there.

The fact that such a precise matching is possible implies that the 
eigenvalue spectrum is completely determined by $N$, the size of 
the loop, and its trace. Hence the eigenvalues are not really 
independent degrees of freedom, which is unexpected.
Moreover we find the spectra of Wilson loops that are $2\times 2$ 
and larger can also be matched with each other, so the
size of the Wilson loop is not really an extra variable here.
Finally, the $N$ dependence is weak.

We note that our results at this stage rely 
on a comparison that
is visual and impressionistic. We intend to provide
a more quantitative and accurate analysis elsewhere~\cite{BurTepVai}.

We have also investigated the eigenvalue spectra of Polyakov loops. We found that it is always possible to match the Polyakov loop
eigenvalue spectra to those of Wilson loops in 1+1 dimensions
(and hence also to Wilson loops in 2+1 dimensions) by choosing
couplings at which the trace of the Polyakov loop equals that
of the Wilson loop.

The existence of a gap in the eigenvalue spectrum at
weak coupling has a rather general origin in terms of Random
Matrix Theory. On the other hand we know that in a confining theory
$\langle u_w \rangle \stackrel{A\to\infty}{\longrightarrow} 0$
which requires a nearly flat eigenvalue spectrum.
Thus as we decrease the lattice spacing, the eigenvalue spectrum
of a $L\times L$ Wilson loop must change from being
nearly uniform to eventually having a gap.
These considerations do not explain the complete matching of eigenvalue
spectra across space-time dimension and loop size by merely matching traces.

For the gap formation to be physically significant, it must occur at
a fixed physical area in the continuum limit.
However, this is not the case in 2+1 dimensions.
The reason is the
perturbative self-energy of the sources whose propagators
are the straight-line sections of the Wilson loop. (Often
referred to as the `perimeter term'.) Due to these the critical
area for gap formation will vanish as we approach the continuum limit:
\begin{equation}
A_{crit} \propto \frac{1}{(\log\lambda)^2}
\stackrel{a\to 0}{\longrightarrow} 0
\label{eqn_WLr25}
\end{equation}
One can imagine
regularising the divergent self-energies so that  $A_{crit}$
is finite and non-zero in the continuum limit, but then it would 
appear to depend on the regularisation mass scale $\mu$ used.

\section{Conclusions}
We find a very close match between the behaviour of several observables across the bulk transition in $2+1$ dimensions and the Gross--Witten transition in $1+1$ dimensions. In particular the local contribution to the third derivative of the partition function has a very similar discontinuity in both cases. We find also a very close match in the plaquette eigenvalue spectra.

However, there is clearly more than this going on. We see a peak in the specific heat in $2+1$ dimensions that is mainly due to correlations between nearby plaquettes and that grows with $N$, which is not present in $1+1$ dimensions. This suggests that there is either a second--order phase transition at $N=\infty$, or a third--order phase transition that has a critical exponent different to $-1$. The correlations of plaquettes contributing to the specific heat peak behave as if due to monopole condensation, suggesting a connection to the bulk transition in $3+1$ dimensions, which can be understood in terms of the condensation of $Z_N$ monopoles. Thus the bulk transition in $2+1$ dimensions appears to have features in common with both the $D=1+1$ and $D=3+1$ transitions.

We have analysed the behaviour of the eigenvalue spectra of Wilson loops in $2+1$ dimensions. We find a very good match with the spectra of Wilson loops in $1+1$ dimensions. This match appears to hold for loops of all sizes, for all $N$, and at all values of the coupling, as long as the traces of the loops are matched. Furthermore, the matching also works for Polyakov loops. This surprising results implies that the eigenvalues are not really independent degrees of freedom. As a corrolary, it immediately follows that in $D=2+1$ at 
$N=\infty$ a gap will form in the eigenvalue spectrum of a 
Wilson loop at a critical coupling that depends on the
size of the loop. However, the physical consequences of this non--analyticity are unclear since the partition function remains analytic at the non--analyticity, and it occurs at zero physical area in the continuum limit.

\begin	{figure}[p]
\begin	{center}
\leavevmode
\scalebox{0.7}{
\begingroup%
  \makeatletter%
  \newcommand{\GNUPLOTspecial}{%
    \@sanitize\catcode`\%=14\relax\special}%
  \setlength{\unitlength}{0.1bp}%
{\GNUPLOTspecial{!
/gnudict 256 dict def
gnudict begin
/Color false def
/Solid false def
/gnulinewidth 5.000 def
/userlinewidth gnulinewidth def
/vshift -33 def
/dl {10 mul} def
/hpt_ 31.5 def
/vpt_ 31.5 def
/hpt hpt_ def
/vpt vpt_ def
/M {moveto} bind def
/L {lineto} bind def
/R {rmoveto} bind def
/V {rlineto} bind def
/vpt2 vpt 2 mul def
/hpt2 hpt 2 mul def
/Lshow { currentpoint stroke M
  0 vshift R show } def
/Rshow { currentpoint stroke M
  dup stringwidth pop neg vshift R show } def
/Cshow { currentpoint stroke M
  dup stringwidth pop -2 div vshift R show } def
/UP { dup vpt_ mul /vpt exch def hpt_ mul /hpt exch def
  /hpt2 hpt 2 mul def /vpt2 vpt 2 mul def } def
/DL { Color {setrgbcolor Solid {pop []} if 0 setdash }
 {pop pop pop Solid {pop []} if 0 setdash} ifelse } def
/BL { stroke userlinewidth 2 mul setlinewidth } def
/AL { stroke userlinewidth 2 div setlinewidth } def
/UL { dup gnulinewidth mul /userlinewidth exch def
      10 mul /udl exch def } def
/PL { stroke userlinewidth setlinewidth } def
/LTb { BL [] 0 0 0 DL } def
/LTa { AL [1 udl mul 2 udl mul] 0 setdash 0 0 0 setrgbcolor } def
/LT0 { PL [] 1 0 0 DL } def
/LT1 { PL [4 dl 2 dl] 0 1 0 DL } def
/LT2 { PL [2 dl 3 dl] 0 0 1 DL } def
/LT3 { PL [1 dl 1.5 dl] 1 0 1 DL } def
/LT4 { PL [5 dl 2 dl 1 dl 2 dl] 0 1 1 DL } def
/LT5 { PL [4 dl 3 dl 1 dl 3 dl] 1 1 0 DL } def
/LT6 { PL [2 dl 2 dl 2 dl 4 dl] 0 0 0 DL } def
/LT7 { PL [2 dl 2 dl 2 dl 2 dl 2 dl 4 dl] 1 0.3 0 DL } def
/LT8 { PL [2 dl 2 dl 2 dl 2 dl 2 dl 2 dl 2 dl 4 dl] 0.5 0.5 0.5 DL } def
/Pnt { stroke [] 0 setdash
   gsave 1 setlinecap M 0 0 V stroke grestore } def
/Dia { stroke [] 0 setdash 2 copy vpt add M
  hpt neg vpt neg V hpt vpt neg V
  hpt vpt V hpt neg vpt V closepath stroke
  Pnt } def
/Pls { stroke [] 0 setdash vpt sub M 0 vpt2 V
  currentpoint stroke M
  hpt neg vpt neg R hpt2 0 V stroke
  } def
/Box { stroke [] 0 setdash 2 copy exch hpt sub exch vpt add M
  0 vpt2 neg V hpt2 0 V 0 vpt2 V
  hpt2 neg 0 V closepath stroke
  Pnt } def
/Crs { stroke [] 0 setdash exch hpt sub exch vpt add M
  hpt2 vpt2 neg V currentpoint stroke M
  hpt2 neg 0 R hpt2 vpt2 V stroke } def
/TriU { stroke [] 0 setdash 2 copy vpt 1.12 mul add M
  hpt neg vpt -1.62 mul V
  hpt 2 mul 0 V
  hpt neg vpt 1.62 mul V closepath stroke
  Pnt  } def
/Star { 2 copy Pls Crs } def
/BoxF { stroke [] 0 setdash exch hpt sub exch vpt add M
  0 vpt2 neg V  hpt2 0 V  0 vpt2 V
  hpt2 neg 0 V  closepath fill } def
/TriUF { stroke [] 0 setdash vpt 1.12 mul add M
  hpt neg vpt -1.62 mul V
  hpt 2 mul 0 V
  hpt neg vpt 1.62 mul V closepath fill } def
/TriD { stroke [] 0 setdash 2 copy vpt 1.12 mul sub M
  hpt neg vpt 1.62 mul V
  hpt 2 mul 0 V
  hpt neg vpt -1.62 mul V closepath stroke
  Pnt  } def
/TriDF { stroke [] 0 setdash vpt 1.12 mul sub M
  hpt neg vpt 1.62 mul V
  hpt 2 mul 0 V
  hpt neg vpt -1.62 mul V closepath fill} def
/DiaF { stroke [] 0 setdash vpt add M
  hpt neg vpt neg V hpt vpt neg V
  hpt vpt V hpt neg vpt V closepath fill } def
/Pent { stroke [] 0 setdash 2 copy gsave
  translate 0 hpt M 4 {72 rotate 0 hpt L} repeat
  closepath stroke grestore Pnt } def
/PentF { stroke [] 0 setdash gsave
  translate 0 hpt M 4 {72 rotate 0 hpt L} repeat
  closepath fill grestore } def
/Circle { stroke [] 0 setdash 2 copy
  hpt 0 360 arc stroke Pnt } def
/CircleF { stroke [] 0 setdash hpt 0 360 arc fill } def
/C0 { BL [] 0 setdash 2 copy moveto vpt 90 450  arc } bind def
/C1 { BL [] 0 setdash 2 copy        moveto
       2 copy  vpt 0 90 arc closepath fill
               vpt 0 360 arc closepath } bind def
/C2 { BL [] 0 setdash 2 copy moveto
       2 copy  vpt 90 180 arc closepath fill
               vpt 0 360 arc closepath } bind def
/C3 { BL [] 0 setdash 2 copy moveto
       2 copy  vpt 0 180 arc closepath fill
               vpt 0 360 arc closepath } bind def
/C4 { BL [] 0 setdash 2 copy moveto
       2 copy  vpt 180 270 arc closepath fill
               vpt 0 360 arc closepath } bind def
/C5 { BL [] 0 setdash 2 copy moveto
       2 copy  vpt 0 90 arc
       2 copy moveto
       2 copy  vpt 180 270 arc closepath fill
               vpt 0 360 arc } bind def
/C6 { BL [] 0 setdash 2 copy moveto
      2 copy  vpt 90 270 arc closepath fill
              vpt 0 360 arc closepath } bind def
/C7 { BL [] 0 setdash 2 copy moveto
      2 copy  vpt 0 270 arc closepath fill
              vpt 0 360 arc closepath } bind def
/C8 { BL [] 0 setdash 2 copy moveto
      2 copy vpt 270 360 arc closepath fill
              vpt 0 360 arc closepath } bind def
/C9 { BL [] 0 setdash 2 copy moveto
      2 copy  vpt 270 450 arc closepath fill
              vpt 0 360 arc closepath } bind def
/C10 { BL [] 0 setdash 2 copy 2 copy moveto vpt 270 360 arc closepath fill
       2 copy moveto
       2 copy vpt 90 180 arc closepath fill
               vpt 0 360 arc closepath } bind def
/C11 { BL [] 0 setdash 2 copy moveto
       2 copy  vpt 0 180 arc closepath fill
       2 copy moveto
       2 copy  vpt 270 360 arc closepath fill
               vpt 0 360 arc closepath } bind def
/C12 { BL [] 0 setdash 2 copy moveto
       2 copy  vpt 180 360 arc closepath fill
               vpt 0 360 arc closepath } bind def
/C13 { BL [] 0 setdash  2 copy moveto
       2 copy  vpt 0 90 arc closepath fill
       2 copy moveto
       2 copy  vpt 180 360 arc closepath fill
               vpt 0 360 arc closepath } bind def
/C14 { BL [] 0 setdash 2 copy moveto
       2 copy  vpt 90 360 arc closepath fill
               vpt 0 360 arc } bind def
/C15 { BL [] 0 setdash 2 copy vpt 0 360 arc closepath fill
               vpt 0 360 arc closepath } bind def
/Rec   { newpath 4 2 roll moveto 1 index 0 rlineto 0 exch rlineto
       neg 0 rlineto closepath } bind def
/Square { dup Rec } bind def
/Bsquare { vpt sub exch vpt sub exch vpt2 Square } bind def
/S0 { BL [] 0 setdash 2 copy moveto 0 vpt rlineto BL Bsquare } bind def
/S1 { BL [] 0 setdash 2 copy vpt Square fill Bsquare } bind def
/S2 { BL [] 0 setdash 2 copy exch vpt sub exch vpt Square fill Bsquare } bind def
/S3 { BL [] 0 setdash 2 copy exch vpt sub exch vpt2 vpt Rec fill Bsquare } bind def
/S4 { BL [] 0 setdash 2 copy exch vpt sub exch vpt sub vpt Square fill Bsquare } bind def
/S5 { BL [] 0 setdash 2 copy 2 copy vpt Square fill
       exch vpt sub exch vpt sub vpt Square fill Bsquare } bind def
/S6 { BL [] 0 setdash 2 copy exch vpt sub exch vpt sub vpt vpt2 Rec fill Bsquare } bind def
/S7 { BL [] 0 setdash 2 copy exch vpt sub exch vpt sub vpt vpt2 Rec fill
       2 copy vpt Square fill
       Bsquare } bind def
/S8 { BL [] 0 setdash 2 copy vpt sub vpt Square fill Bsquare } bind def
/S9 { BL [] 0 setdash 2 copy vpt sub vpt vpt2 Rec fill Bsquare } bind def
/S10 { BL [] 0 setdash 2 copy vpt sub vpt Square fill 2 copy exch vpt sub exch vpt Square fill
       Bsquare } bind def
/S11 { BL [] 0 setdash 2 copy vpt sub vpt Square fill 2 copy exch vpt sub exch vpt2 vpt Rec fill
       Bsquare } bind def
/S12 { BL [] 0 setdash 2 copy exch vpt sub exch vpt sub vpt2 vpt Rec fill Bsquare } bind def
/S13 { BL [] 0 setdash 2 copy exch vpt sub exch vpt sub vpt2 vpt Rec fill
       2 copy vpt Square fill Bsquare } bind def
/S14 { BL [] 0 setdash 2 copy exch vpt sub exch vpt sub vpt2 vpt Rec fill
       2 copy exch vpt sub exch vpt Square fill Bsquare } bind def
/S15 { BL [] 0 setdash 2 copy Bsquare fill Bsquare } bind def
/D0 { gsave translate 45 rotate 0 0 S0 stroke grestore } bind def
/D1 { gsave translate 45 rotate 0 0 S1 stroke grestore } bind def
/D2 { gsave translate 45 rotate 0 0 S2 stroke grestore } bind def
/D3 { gsave translate 45 rotate 0 0 S3 stroke grestore } bind def
/D4 { gsave translate 45 rotate 0 0 S4 stroke grestore } bind def
/D5 { gsave translate 45 rotate 0 0 S5 stroke grestore } bind def
/D6 { gsave translate 45 rotate 0 0 S6 stroke grestore } bind def
/D7 { gsave translate 45 rotate 0 0 S7 stroke grestore } bind def
/D8 { gsave translate 45 rotate 0 0 S8 stroke grestore } bind def
/D9 { gsave translate 45 rotate 0 0 S9 stroke grestore } bind def
/D10 { gsave translate 45 rotate 0 0 S10 stroke grestore } bind def
/D11 { gsave translate 45 rotate 0 0 S11 stroke grestore } bind def
/D12 { gsave translate 45 rotate 0 0 S12 stroke grestore } bind def
/D13 { gsave translate 45 rotate 0 0 S13 stroke grestore } bind def
/D14 { gsave translate 45 rotate 0 0 S14 stroke grestore } bind def
/D15 { gsave translate 45 rotate 0 0 S15 stroke grestore } bind def
/DiaE { stroke [] 0 setdash vpt add M
  hpt neg vpt neg V hpt vpt neg V
  hpt vpt V hpt neg vpt V closepath stroke } def
/BoxE { stroke [] 0 setdash exch hpt sub exch vpt add M
  0 vpt2 neg V hpt2 0 V 0 vpt2 V
  hpt2 neg 0 V closepath stroke } def
/TriUE { stroke [] 0 setdash vpt 1.12 mul add M
  hpt neg vpt -1.62 mul V
  hpt 2 mul 0 V
  hpt neg vpt 1.62 mul V closepath stroke } def
/TriDE { stroke [] 0 setdash vpt 1.12 mul sub M
  hpt neg vpt 1.62 mul V
  hpt 2 mul 0 V
  hpt neg vpt -1.62 mul V closepath stroke } def
/PentE { stroke [] 0 setdash gsave
  translate 0 hpt M 4 {72 rotate 0 hpt L} repeat
  closepath stroke grestore } def
/CircE { stroke [] 0 setdash 
  hpt 0 360 arc stroke } def
/Opaque { gsave closepath 1 setgray fill grestore 0 setgray closepath } def
/DiaW { stroke [] 0 setdash vpt add M
  hpt neg vpt neg V hpt vpt neg V
  hpt vpt V hpt neg vpt V Opaque stroke } def
/BoxW { stroke [] 0 setdash exch hpt sub exch vpt add M
  0 vpt2 neg V hpt2 0 V 0 vpt2 V
  hpt2 neg 0 V Opaque stroke } def
/TriUW { stroke [] 0 setdash vpt 1.12 mul add M
  hpt neg vpt -1.62 mul V
  hpt 2 mul 0 V
  hpt neg vpt 1.62 mul V Opaque stroke } def
/TriDW { stroke [] 0 setdash vpt 1.12 mul sub M
  hpt neg vpt 1.62 mul V
  hpt 2 mul 0 V
  hpt neg vpt -1.62 mul V Opaque stroke } def
/PentW { stroke [] 0 setdash gsave
  translate 0 hpt M 4 {72 rotate 0 hpt L} repeat
  Opaque stroke grestore } def
/CircW { stroke [] 0 setdash 
  hpt 0 360 arc Opaque stroke } def
/BoxFill { gsave Rec 1 setgray fill grestore } def
end
}}%
\begin{picture}(3600,2160)(0,0)%
{\GNUPLOTspecial{"
gnudict begin
gsave
0 0 translate
0.100 0.100 scale
0 setgray
newpath
1.000 UL
LTb
450 300 M
63 0 V
2937 0 R
-63 0 V
450 520 M
63 0 V
2937 0 R
-63 0 V
450 740 M
63 0 V
2937 0 R
-63 0 V
450 960 M
63 0 V
2937 0 R
-63 0 V
450 1180 M
63 0 V
2937 0 R
-63 0 V
450 1400 M
63 0 V
2937 0 R
-63 0 V
450 1620 M
63 0 V
2937 0 R
-63 0 V
450 1840 M
63 0 V
2937 0 R
-63 0 V
450 2060 M
63 0 V
2937 0 R
-63 0 V
450 300 M
0 63 V
0 1697 R
0 -63 V
825 300 M
0 63 V
0 1697 R
0 -63 V
1200 300 M
0 63 V
0 1697 R
0 -63 V
1575 300 M
0 63 V
0 1697 R
0 -63 V
1950 300 M
0 63 V
0 1697 R
0 -63 V
2325 300 M
0 63 V
0 1697 R
0 -63 V
2700 300 M
0 63 V
0 1697 R
0 -63 V
3075 300 M
0 63 V
0 1697 R
0 -63 V
3450 300 M
0 63 V
0 1697 R
0 -63 V
1.000 UL
LTb
450 300 M
3000 0 V
0 1760 V
-3000 0 V
450 300 L
1.000 UL
LT0
450 418 M
209 52 V
520 399 V
261 288 V
130 93 V
130 10 V
261 -158 V
2221 914 L
2742 692 L
3263 582 L
187 -23 V
1.000 UL
LT1
1179 621 M
261 321 V
260 498 V
130 17 V
131 -127 V
2221 982 L
2742 718 L
3263 595 L
187 -24 V
1.000 UL
LT2
1179 612 M
261 176 V
260 480 V
130 383 V
65 -27 V
66 -189 V
2221 995 L
2481 832 L
2742 718 L
3263 617 L
187 -32 V
1.000 UL
LT3
1440 775 M
260 449 V
65 176 V
65 295 V
33 105 V
32 -114 V
66 -229 V
2221 995 L
0.600 UP
1.000 UL
LT4
659 469 M
0 2 V
-31 -2 R
62 0 V
-62 2 R
62 0 V
489 397 R
0 2 V
-31 -2 R
62 0 V
-62 2 R
62 0 V
230 286 R
0 2 V
-31 -2 R
62 0 V
-62 2 R
62 0 V
99 88 R
0 9 V
-31 -9 R
62 0 V
-62 9 R
62 0 V
99 4 R
0 2 V
-31 -2 R
62 0 V
-62 2 R
62 0 V
230 -160 R
0 2 V
-31 -2 R
62 0 V
-62 2 R
62 0 V
2221 913 M
0 1 V
-31 -1 R
62 0 V
-62 1 R
62 0 V
2742 683 M
0 17 V
-31 -17 R
62 0 V
-62 17 R
62 0 V
3263 573 M
0 17 V
-31 -17 R
62 0 V
-62 17 R
62 0 V
659 470 Pls
1179 869 Pls
1440 1157 Pls
1570 1250 Pls
1700 1260 Pls
1961 1102 Pls
2221 914 Pls
2742 692 Pls
3263 582 Pls
0.600 UP
1.000 UL
LT5
1179 612 M
0 18 V
-31 -18 R
62 0 V
-62 18 R
62 0 V
230 295 R
0 35 V
-31 -35 R
62 0 V
-62 35 R
62 0 V
229 462 R
0 35 V
-31 -35 R
62 0 V
-62 35 R
62 0 V
99 -22 R
0 44 V
-31 -44 R
62 0 V
-62 44 R
62 0 V
100 -158 R
0 17 V
-31 -17 R
62 0 V
-62 17 R
62 0 V
2221 973 M
0 18 V
-31 -18 R
62 0 V
-62 18 R
62 0 V
2742 714 M
0 8 V
-31 -8 R
62 0 V
-62 8 R
62 0 V
3263 590 M
0 9 V
-31 -9 R
62 0 V
-62 9 R
62 0 V
1179 621 Crs
1440 942 Crs
1700 1440 Crs
1830 1457 Crs
1961 1330 Crs
2221 982 Crs
2742 718 Crs
3263 595 Crs
0.600 UP
1.000 UL
LT6
1179 590 M
0 44 V
-31 -44 R
62 0 V
-62 44 R
62 0 V
230 128 R
0 53 V
-31 -53 R
62 0 V
-62 53 R
62 0 V
229 431 R
0 44 V
-31 -44 R
62 0 V
-62 44 R
62 0 V
99 343 R
0 35 V
-31 -35 R
62 0 V
-62 35 R
62 0 V
34 -88 R
0 88 V
-31 -88 R
62 0 V
-62 88 R
62 0 V
35 -255 R
0 44 V
-31 -44 R
62 0 V
-62 44 R
62 0 V
2221 982 M
0 26 V
-31 -26 R
62 0 V
-62 26 R
62 0 V
2481 824 M
0 17 V
-31 -17 R
62 0 V
-62 17 R
62 0 V
2742 709 M
0 18 V
-31 -18 R
62 0 V
-62 18 R
62 0 V
3263 612 M
0 9 V
-31 -9 R
62 0 V
-62 9 R
62 0 V
1179 612 Star
1440 788 Star
1700 1268 Star
1830 1651 Star
1895 1624 Star
1961 1435 Star
2221 995 Star
2481 832 Star
2742 718 Star
3263 617 Star
0.600 UP
1.000 UL
LT7
1440 749 M
0 53 V
-31 -53 R
62 0 V
-62 53 R
62 0 V
229 391 R
0 62 V
-31 -62 R
62 0 V
-62 62 R
62 0 V
34 105 R
0 80 V
-31 -80 R
62 0 V
-62 80 R
62 0 V
34 224 R
0 62 V
-31 -62 R
62 0 V
-62 62 R
62 0 V
2 26 R
0 97 V
-31 -97 R
62 0 V
-62 97 R
62 0 V
1 -220 R
0 114 V
-31 -114 R
62 0 V
-62 114 R
62 0 V
35 -321 R
0 70 V
-31 -70 R
62 0 V
-62 70 R
62 0 V
2221 978 M
0 35 V
-31 -35 R
62 0 V
-62 35 R
62 0 V
1440 775 Box
1700 1224 Box
1765 1400 Box
1830 1695 Box
1863 1800 Box
1895 1686 Box
1961 1457 Box
2221 995 Box
stroke
grestore
end
showpage
}}%
\put(1950,50){\makebox(0,0){$\gamma$}}%
\put(100,1180){%
\makebox(0,0)[b]{\shortstack{$C_o$}}%
}%
\put(3450,200){\makebox(0,0){0.65}}%
\put(3075,200){\makebox(0,0){0.6}}%
\put(2700,200){\makebox(0,0){0.55}}%
\put(2325,200){\makebox(0,0){0.5}}%
\put(1950,200){\makebox(0,0){0.45}}%
\put(1575,200){\makebox(0,0){0.4}}%
\put(1200,200){\makebox(0,0){0.35}}%
\put(825,200){\makebox(0,0){0.3}}%
\put(450,200){\makebox(0,0){0.25}}%
\put(400,2060){\makebox(0,0)[r]{0.4}}%
\put(400,1840){\makebox(0,0)[r]{0.35}}%
\put(400,1620){\makebox(0,0)[r]{0.3}}%
\put(400,1400){\makebox(0,0)[r]{0.25}}%
\put(400,1180){\makebox(0,0)[r]{0.2}}%
\put(400,960){\makebox(0,0)[r]{0.15}}%
\put(400,740){\makebox(0,0)[r]{0.1}}%
\put(400,520){\makebox(0,0)[r]{0.05}}%
\put(400,300){\makebox(0,0)[r]{0}}%
\end{picture}%
\endgroup
 
}
\end	{center}
\caption{The plaquette correlator, $C_o$, as a function of 
$\gamma=\frac{\beta}{2N^2}$
for SU(6) ($+$), SU(12) ($\times$), 
SU(24) ($\ast$) and SU(48) ($\Box$).}  
\label{fig_outplane}
\end 	{figure}

\end{document}